\documentclass[prab, reprint, superscriptaddress]{revtex4-2}
\usepackage{graphicx, setspace}
\usepackage{booktabs}
\usepackage{amsmath}  % foundation of ams-latex
\usepackage{amssymb}  % new fonts and names for them
\usepackage{mathrsfs}    % required package for font scripting
\usepackage{euscript}     % extra Euler font 
\usepackage{eepic}
\usepackage{verbatim}
\usepackage{physics, siunitx, soul, color}

\graphicspath{{./pics/}}

\begin{document}

\title{Transverse Gradient Undulator in a Storage Ring X-ray Free Electron Laser Oscillator}
\date{\today}

\author{Yuanshen \surname{Li}}
\email{ysli@uchicago.edu}
\affiliation{University of Chicago, Chicago, Illinois 60637, USA}
\author{Ryan \surname{Lindberg}}
\affiliation{Argonne National Laboratory, Argonne, Illinois 60439, USA}
\author{Kwang-Je \surname{Kim}}
\affiliation{University of Chicago, Chicago, Illinois 60637, USA}
\affiliation{Argonne National Laboratory, Argonne, Illinois 60439, USA}

\begin{abstract}
Modern electron storage rings produce bright X-rays via spontaneous synchrotron emission, which is useful for a variety of scientific applications. The X-ray free-electron laser oscillator (XFELO) has the potential to amplify this output, both in terms of peak power and photon coherence. However, even current 4th generation storage rings (4GSRs) lack the requisite electron beam brightness to drive the XFELO due to its electron energy spread. The transverse gradient undulator (TGU) can overcome this issue, thus providing a practical means to couple the 4GSR to the XFELO. In this study, we first examine the theoretical basis of the TGU interaction by deriving the 3D TGU gain formula in the low gain approximation. Then, we perform an optimization study of the gain formula in order to determine optimal beam and machine parameters. Finally, we construct a hypothetical storage ring TGU-XFELO based on near-optimal parameters and report on its projected performance using multi-stage numerical simulation. We also discuss potential implementation challenges associated with the ring-FEL coupling.
\end{abstract}

\maketitle

\section{\label{sec:intro}Introduction}
Synchrotron emission from high-energy electrons can provide intense radiation over a wide spectral range that extends to hard X-rays. Modern third generation electron storage rings (3GSRs) utilize this spontaneous emission to produce bright X-rays that are used to measure and probe a host of materials, systems, and processes. Going one step further, recent advances in self-amplified spontaneous emission (SASE) free-electron lasers (FELs) offer yet more orders of magnitude improvement in peak brightness and optical coherence~\cite{lclspaper}.

However, unlike conventional lasers, SASE FELs do not offer full temporal coherence. One method to remedy this would be to use an oscillator, in which the photon beam is trapped within a high-efficiency optical cavity.  This FEL oscillator setup has been used to great success in the infrared (IR) through ultraviolent (UV) wavelengths~\cite{orsayfel, dukefel}. The electron driver has typically been a linear accelerator (linac) for IR and a synchrotron for UV. 

In principle, the FEL oscillator setup is extensible to the X-ray regime (wavelength $\lesssim 3\, \si{\angstrom}$) by employing Bragg reflectors as cavity mirrors~\cite{xraycavity, xraycavity2}. However, the electron beam brightness in current 3GSRs is not sufficient to produce the necessary FEL gain --- the natural emittance $\varepsilon_{x,0}$ is typically one or two orders of magnitude too large, while energy spread $\sigma_{\gamma}$ exceeds the ideal by a factor of ten.

Recent advances in storage ring technology have managed to narrow this gap. The nascent fourth generation storage ring (4GSR) uses multi-bend achromats to drastically reduce transverse natural emittance to $\sim 10^{-11}$ m~rad, thus fulfilling the requirement that $\varepsilon_{x,0} \lesssim \lambda/4\pi$ for wavelengths in the hard X-ray range~\cite{petra4, thebook}. However, electron energy spread remains too large --- typically having a value of $\sigma_{\gamma}/\gamma \sim 0.1\%$ at equilibrium.

\begin{figure}
 \includegraphics[width=0.6\linewidth]{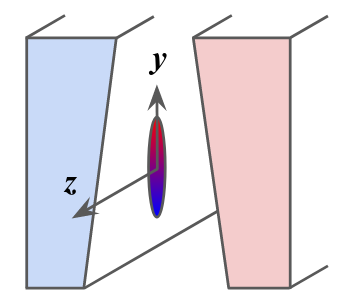}
 \caption{\label{fig:TGUdemo}Cross-section of a transverse gradient undulator (TGU) aligned vertically. Direction of electron motion is in $z$. The magnetic field gradient grows stronger along the positive $y$-axis due to the canted undulator poles (exaggerated for visibility). If electron bunch (center oval) is dispersed appropriately so that higher energy electrons (red; top half) experience higher field strength while lower energy electrons (blue; bottom half) experience less, then FEL gain can be greatly improved even for large electron energy spreads.}
 \end{figure}

One method to circumvent the energy spread limitation is to use a transverse gradient undulator (TGU)~\cite{tguoriginal, tgu2013, agapov2018}. In contrast to a planar undulator, the magnetic pole faces of the TGU is slightly canted along one transverse axis (Fig.~\ref{fig:TGUdemo}). To first order, the undulator field strength becomes a transversely linear function. If the electron beam energy is appropriately correlated along the same axis, one can achieve significant FEL gain even with large energy spreads. The TGU concept was recently revisited for high-gain FELs driven by large energy spread beams produced by laser-plasma accelerators~\cite{tgulaserplasma}, and for high-gain FELs in an electron synchrotron~\cite{tguhighgain}. 

\begin{figure*}
  \includegraphics[width=0.75\linewidth]{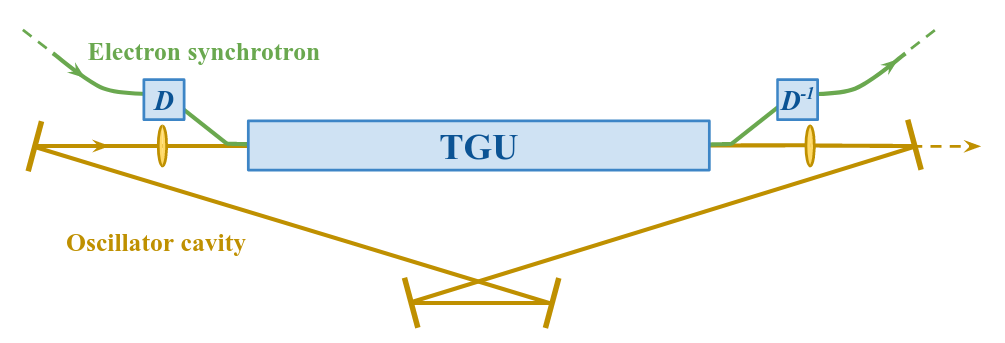}
  \caption{\label{fig:srxfelo}Schematic of the TGU-XFELO driven by an electron synchrotron. The three constituent components are the electron synchrotron (green), the TGU mechanism (blue), and the optical cavity (orange). The TGU mechanism consists of the undulator itself, as well as dispersive sections denoted by $D$ and $D^{-1}$ above.}
\end{figure*}

 In this paper, we will examine the application of the TGU to a low-gain X-ray oscillator driven by an electron synchrotron (Fig.~\ref{fig:srxfelo}). We begin by discussing low-gain TGU theory, culminating in the derivation of the 3D gain formula in Section~\ref{sec:lowgaintheory}. Next, in section~\ref{sec:tguoptimization} we perform parameter optimization on the 3D gain formula in order to determine beam and machine parameters for maximum gain. Finally, in section~\ref{sec:srxfelo}, we report on the projected performance of a hypothetical storage ring TGU-XFELO based on multi-stage numerical simulation. We will also examine the relevant storage ring FEL dynamics and discuss practical implementation concerns.

While this study focuses exclusively on storage rings as the electron driver, we should note that linac driven FELs are also a hugely successful and rapidly advancing subfield of FEL technology. In contrast to the storage ring, the linac is able to achieve much higher electron beam brightness, which, in conjunction with the advent of superconducting radiofrequency (SCRF) cavity technology, allows it to become a serious contender in driving the low-gain X-ray oscillator~\cite{euxfel}.  However, one major drawback of the linac design is its low user capacity --- in a storage ring, the XFELO would potentially exist alongside many other insertion devices and user stations.
% The use of superconducting technology also drives up construction and operation complexity.
Ultimately, we view these two approaches to light source design as complementary, rather than competitive.

\section{\label{sec:lowgaintheory}Low Gain TGU Theory}
For a planar undulator, the FEL resonance condition stipulates that
\begin{equation}
  \lambda_1 = \lambda_u \frac{1+K^2/2}{2\gamma^2}
  \label{eq:felres}
\end{equation}
where $\lambda_1$ is the resonant radiation wavelength, $\lambda_u$ is the undulator period, $K \equiv eB/(mc k_u)$ is the undulator deflection parameter, and $\gamma$ is the electron Lorentz factor. Additionally, $B$ is the peak undulator magnetic field, $k_u \equiv 2\pi/\lambda_u$ is the undulator wavenumber, and $e$, $m$, $c$ are the electron charge magnitude, mass, and speed of light respectively.

To achieve FEL amplification, the spread in $\lambda_1$ has to be narrower than the FEL gain bandwidth, which in the low gain oscillator is determined by $1/N_u$, where $N_u$ is the number of undulator periods. The spread in $\lambda_1$ is in turn set by the electron energy spread $\sigma_{\gamma}$ via Eq.~\eqref{eq:felres}. Hence, for a fixed $K$ and $\lambda_u$, we require $\sigma_{\gamma}/\gamma \ll 1/N_u$. For large undulators, this limit is typically $\sim 10^{-3}$.

The energy spread requirement can be significantly relaxed if we are able to vary $K$ in the numerator so as to cancel the spread in $\gamma$ in the denominator. The transverse gradient undulator (TGU) accomplishes this using a two-step method. First, by slightly canting the pole faces along the $y$-axis (Fig.~\ref{fig:TGUdemo}), the on-axis field strength becomes an approximately linear function of $y$:
\begin{equation}
  K(y) \approx K_0(1+\alpha y),
  \label{eq:tgugradientlinear}
\end{equation}
where $\alpha$ is the TGU gradient parameter. Second, we introduce a correlation between the electrons' energy and their transverse position via dispersion, such that 
\begin{equation}
  y_j = D\eta_j + {y_\beta}_j,
  \label{eq:dispersion}
\end{equation}
where $y_j$ is the position of the $j$th electron, $D$ is the dispersion parameter, $\eta_j \equiv (\gamma_j-\gamma_0)/\gamma_0$ is the relative deviation from the nominal electron energy $\gamma_0$, and ${y_{\beta}}_j$ is the original betatron trajectory of the electron. For each electron $j$, we would like equation~\eqref{eq:felres} to be individually satisfied:
\begin{equation}
  \lambda_1 = \lambda_u \frac{1+K_0^2(1+\alpha y_j)^2/2}{2\gamma_0^2(1 + \eta_j)^2}.
  \label{eq:felrestgu}
\end{equation}
Inserting equation~\eqref{eq:dispersion} into the above yields
\begin{equation}
  \lambda_1 \approx \lambda_u \frac{1+K^2_0/2}{2\gamma_0^2}\left[1 + \frac{K_0^2\alpha(D\eta_j + {y_\beta}_j)}{1+K_0^2/2} - 2\eta_j\right],
              \label{eq:tgures1}
\end{equation}
for $\eta_j \ll 1$ and $\alpha y_j \ll 1$. We can eliminate $\eta_j$ and effectively remove the influence of energy spread by choosing
\begin{equation}
  \alpha D = \frac{2 + K_0^2}{K_0^2},
  \label{eq:tgures2}
\end{equation}
and assuming $y_{\beta_j} \ll D\eta_j$. The latter assumption implies that for the overall electron ensemble, the correlated electron beam size in $y$ should be dominated by dispersion, i.e. $\langle {y_\beta}_j^2 \rangle \equiv \sigma_y^2 \ll D^2\sigma_\eta^2$. We quantify this by introducing the dimensionless TGU parameter
\begin{equation}
  \Gamma \equiv \frac{D\sigma_{\eta}}{\sigma_y}.
  \label{eq:tguparam}
\end{equation}
The previous argument is then equivalent to $\Gamma \gg 1$. For $\Gamma \sim 1$ or less, the TGU benefit is greatly diminished. We will see that $\Gamma$ plays a pivotal role in TGU gain.

Previous studies by Kroll \emph{et al.} have found that the TGU also excites transverse betatron oscillations~\cite{KMRtgu}. This effect, however, is only dominant when the electron bunch undergoes many cycles of betatron oscillation within the length of the TGU, which does not hold true in the X-ray regime~\cite{thebook}. In fact, in the latter case, the TGU focusing strength is $\alpha/k_u$ times weaker than natural undulator focusing, which itself is considered small. (Typical values of $\alpha/k_u \sim 0.1$ in our case.) Hence going forward, we will assume the TGU-induced transverse motion to be negligible. This is especially important for a key assumption of the gain convolution formula, to be discussed next.

\subsection{Derivation of 3D gain formula}
The derivation of the full 3D TGU gain formula follows closely the steps of deriving the gain formula for a regular planar undulator, presented in \cite{thebook, kjnima318}, with two key differences. Firstly, we make the substitution
\begin{equation}
  y \rightarrow y-D\eta
\end{equation}
in the particle position coordinate to account for the added TGU dispersion. Secondly, in accordance with Eq.~\eqref{eq:tgures1}, the TGU magnetic gradient modifies the FEL resonance condition:
\begin{equation}
  \frac{k_u}{k_1} \approx \frac{1+K_0^2/2}{2\gamma_0^2} \left[ 1 + \frac{K_0^2\alpha y}{1+K_0^2/2} - 2\eta  \right],
\end{equation}
where $k_u,k_1$ are the wavenumbers of the undulator and the fundamental radiation harmonic respectively. This change is carried forward into the evolution equation of the FEL ponderomotive phase $\theta$:
\begin{equation}
  \dv{\theta}{z} = 2k_u\eta - k_uT_{\alpha}y - \frac{k_1}{2}\left( \va*{p}^2 + k_{\beta}^2\va*{x}^2 \right),
\end{equation}
where $T_{\alpha} \equiv K_0^2\alpha/(1+K_0^2/2)$, $(\va*{x},\va*{p})$ are the particle position and momentum coordinates, and $k_{\beta}$ represents any external focusing experienced by the electron. The first and third terms are identical to that of a planar undulator. Since we are ignoring TGU-induced transverse focusing effects, we leave the third term unchanged, as well as the evolution equations for $\va*{x}$ and $\va*{p}$. For our purposes the TGU magnetic gradient is only important for longitudinal FEL dynamics.

With all of that in mind, the TGU gain formula can be obtained by retracing the derivation in \cite{thebook, kjnima318}. We outline here the general steps, with more details presented in the Appendix. To begin, we define the brightness of a radiation field $R$ using its Wigner transform \cite{kjnima246}:
\begin{equation}
  B_R(\va*{y},\va*{\phi}) = \int d\va*{\xi}\, R^*(\va*{\phi}+ \va*{\xi}/2)R(\va*{\phi}-\va*{\xi}/2)e^{-ik\va*{y}\vdot\va*{\phi}},
  \label{eq:wigner}
\end{equation}
where $(\va*{y},\va*{\phi})$ are the spatial and angular coordinates of the radiation field. The brightness function $B_R$ can be thought of as analogous to the electron phase space distribution function $F(\eta, \va*{x},\va*{p})$. Then FEL gain can be obtained via the convolution of the radiation and particle distributions.
\begin{widetext}
More concretely, 
\begin{equation}
  G = \frac{G_0}{8\pi N_uL_u^2\lambda_1^2} \int d\eta d \va*{x} d \va*{y} d\va*{\phi} d\va*{p} \, B_E(\va*{y}, \va*{\phi}) B_U(\eta, \va*{x} - \va*{y}, \va*{\phi} - \va*{p}) \pdv{\eta} F(\eta, \va*{x}, \va*{p}).
  \label{eq:convformula}
\end{equation}
Here $B_E$ is the seed radiation brightness, $B_U$ is the brightness of the spontaneous radiation, $F$ is phase space distribution of the initial electron beam, and $L_u$ is the undulator length.
\end{widetext}
The prefactor $G_0$ is defined as
\begin{equation}
  G_0 = (4\pi)^2 \gamma_0 \frac{I}{I_A} \frac{K_0^2[JJ]^2}{(1+K_0^2/2)^2} N_u^3\lambda_1^2,
        \label{eq:gainprefactor}
\end{equation}
where $I$ is the electron peak current, $I_A = 4\pi \epsilon_0mc^3/e \approx 17 \, \si{kA}$ is the Alfv\'en current with $\epsilon_0$ being the vacuum permittivity, and the Bessel function factor $[JJ] = J_0(K_0^2/(4+2K_0^2)) - J_1(K_0^2/(4+2K_0^2))$. Note that Eq.~\eqref{eq:convformula} is derived under the assumption that the electrons undergo no transverse focusing in the undulator, i.e. $k_{\beta} \rightarrow 0$. This includes natural undulator focusing, external focusing quadrupoles and TGU-induced focusing.

%The two transverse axes are decoupled, so we will restrict ourselves to the TGU axis $y$ to simplify discussion. This allows us to drop the vector notation. (The $x$-axis result can be easily obtained later by setting $\Gamma \rightarrow 0$.)
\begin{widetext}
First, the spontaneous undulator brightness $B_U$ can be obtained from the Wigner transform of the undulator radiation field $\mathcal{U}$. The latter is given by
\begin{equation}
  \mathcal{U}(\eta, \va*{x}(z), \va*{\phi}- \va*{p}; z) = \int_{-L_u/2}^{L_u/2}dz \, \exp \left[ -ik \va*{\phi} \vdot \va*{x}(z) -ikz ( \va*{\phi}- \va*{p} )^2/2 + ik_uz(2\nu \eta - T_{\alpha}\va*{x}(z) - \Delta\nu ) \right],
  \label{eq:brightnessundulator}
\end{equation}
where $\nu \equiv \lambda_1/\lambda$, $\Delta \nu \equiv \nu-1$, and
\begin{equation}
  \va*{x}(z) = \va*{x} - \va*{p}\left( L_u/2 - z\right)
  \label{eq:electrontrajectoryundulator}
\end{equation}
represents the electron trajectory under the no-focusing assumption. Note that on right hand side $(\va*{x}, \va*{p})$ represent the transverse coordinates of the electron at the midpoint of the TGU, i.e. $z=L_u/2$. We choose to evaluate the gain integral at the midpoint because it is the location of the waists of the electron and X-ray beams. Also notice in Eq.~\eqref{eq:brightnessundulator} the inclusion of the $T_{\alpha}$ term which accounts for the TGU's impact on the longitudinal dynamics. Everything else remains unchanged from the planar undulator case.

At the undulator midpoint, the brightness of the Gaussian seed X-ray pulse can be easily expressed as
\begin{equation}
  B_E(y, \phi_y) = \frac{1}{ 2\pi \sigma_{ry} \sigma_{\phi y} } \exp \left( - \frac{y^2}{2\sigma_{ry}^2}- \frac{\phi_y^2}{2\sigma_{\phi y}^2} \right),
      \label{eq:brightnessradiation}
\end{equation}
where $\sigma_{ry}, \sigma_{\phi y}$ are the root-mean-square (RMS) X-ray beam size and divergence measured at $z=L_u/2$. (We drop the vector notation and focus only on the TGU $y$-axis since the two transverse axes are decoupled.) Similarly, we approximate the initial electron distribution as Gaussian:
\begin{equation}
  F(\eta, y, p_y) = \frac{1}{ (2\pi)^{3/2}\sigma_y\sigma_{\eta}\sigma_{py}} \exp \left[ - \frac{(y-D\eta)^2}{2\sigma_y^2} - \frac{\eta^2}{2\sigma_{\eta}^2} - \frac{p^2}{2\sigma_{py}^2} \right].
                    \label{eq:brightnessparticle}
\end{equation}
Here $\sigma_{\eta}$ is the relative energy spread, and $\sigma_y, \sigma_{py}$ are the electron beam size and divergence respectively. Notice the dispersion modification in the exponent.

Inserting Eqs.~\eqref{eq:brightnessradiation}, \eqref{eq:brightnessparticle}, and the Wigner transform of Eq.~\eqref{eq:brightnessundulator} into the gain convolution formula~\eqref{eq:convformula} allows us to obtain an analytical formula for $G$. The calculation involves several steps of Gaussian integration, with more details in the Appendix. The final result is
\begin{equation}
  \label{eq:gain}
  G = \frac{G_0}{4\pi} \int^{1/2}_{-1/2}dz\, ds\, \frac{i(z-s)}{\sqrt{\mathfrak{D}_x\mathfrak{D}_y}} \exp \Bigg[  -2i\delta(z-s)  - \frac{2 \tilde{\sigma}_{\eta}^2(z-s)^2}{1+\Gamma^2}- \left( \frac{\Gamma}{1+\Gamma^2} \frac{\tilde{\sigma}_{\eta}}{\tilde{\beta}_y}\right)^2 \frac{(z^2 - s^2)^2}{2} \frac{\mathfrak{d}_y}{\mathfrak{D}_y} \Bigg].
\end{equation}
\end{widetext}
The gain formula is expressed in terms of following dimensionless parameters
\begin{eqnarray}
  \delta &=& \pi N_u(\omega- \omega_1)/\omega_1, \label{eq:detuning} \\
  \tilde \sigma_{\eta} &=& 2\pi N_u \sigma_{\eta}, \\
  \tilde \beta_y &=& \beta_y/L_u,
\end{eqnarray}
where $\omega = 2\pi c/\lambda$ is the radiation angular frequency, and $\beta_y$ is the betatron function in $y$. The parameter $\delta$ represents frequency detuning from the nominal resonant frequency $\omega_1\equiv 2\pi c/\lambda_1$. We also introduced the diffraction factors
\begin{eqnarray}
  \mathfrak{D}_{x,y} &=& \Sigma_{x,y}^2 + sz\, L_u^2 \Sigma_{\phi x,y}^2 \nonumber \\ &&  - i L_u (z-s) \left[ \frac{1}{4k_1} + k_1 \Sigma_{\phi x,y}^2\Sigma_{x,y}^2 \right], \\
  \mathfrak{d}_{y} &=& \Sigma_{y}^2 + sz\,  L_u^2 \sigma_{\phi y}^2 \nonumber \\ && - i L_u (z-s) \left[ \frac{1}{4k_1} + k_1 \sigma_{\phi y}^2\Sigma_{y}^2 \right],
\end{eqnarray}
with 
\begin{eqnarray}
  \Sigma_{y}^2 &=& \sigma_{y}^2 + \sigma_{r y}^2 + D^2\sigma_{\eta}^2, \\
  \Sigma_{x}^2 &=& \sigma_{x}^2 + \sigma_{rx}^2, \\
  \Sigma_{\phi x,y}^2 &=& \sigma_{p x,y}^2 + \sigma_{\phi x,y}^2. \label{eq:Sigmaphixy}
\end{eqnarray}
Here, $\sigma_{x,y}, \sigma_{px,py}$ are the electron beam sizes and divergences in $x,y$ respectively, while $\sigma_{rx,ry},\sigma_{\phi x, \phi y}$ are the analogous quantities for the seed X-ray beam. \par

\subsection{Discussion of 3D gain formula}

In the limit $\Gamma \rightarrow 0$, the gain formula~\eqref{eq:gain} reduces correctly to its counterpart for a traditional planar undulator. As in the planar undulator case, the factor $\mathfrak{D}_x\mathfrak{D}_y$ represents the dilution of gain due to 3D diffraction effects. In the exponential, the first term $-2i\delta(z-s)$ represents the effect of frequency detuning and similarly remains unchanged from its non-TGU counterpart. \par

The impact of the TGU is most evident in the second term $-2 \tilde{\sigma}_{\eta}^2(z-s)^2/(1+\Gamma^2)$. Without the TGU ($\Gamma = 0$), this term results in the exponential suppression of gain due to energy spread. The TGU parameter $\Gamma$ acts in the denominator to mitigate this effect. The third and final term in the exponential in Eq.~\eqref{eq:gain} limits gain when a large electron divergence outweighs the required $y$-$\gamma$ correlation for ideal TGU cancellation. This means that gain does not increase monotonically with $\Gamma$, but rather, reaches a maximum and then falls off.

While not appearing directly in the gain formula, the electron transverse emittances $\varepsilon_{x,y}$ play an important role due to their influence on $\sigma_{x,y}$ and $\sigma_{px,py}$ and thus the diffraction dilution factors $\mathfrak{D}_{x,y}$. In particular, having a small emittance $\varepsilon_y$ along the TGU axis can potentially greatly increase gain. Herein lies the advantage of driving the TGU with a storage ring --- due to the unique nature of radiation damping in a ring, the equilibrium natural emittance $\varepsilon_{x,0} \equiv \varepsilon_x + \varepsilon_y$ is a conserved quantity, with the vertical emittance $\varepsilon_y$ being typically orders of magnitude smaller than its horizontal counterpart. Moreover, by using coupling lattice elements such as skew quadrupoles, we can fine-tune this emittance ratio $k_c \equiv \varepsilon_y / \varepsilon_x$. We will see in the following section that by choosing $k_c \ll 1$, we can potentially increase gain by an order of magnitude. Even modest values of $k_c$, such as $1/6$ used for PETRA-IV~\cite{petra4}, present significant gain improvement (up to 2x).

\section{\label{sec:tguoptimization}TGU Gain Optimization}

For a given set of machine parameters, one may be interested in the optimal beam parameters that result in the highest TGU gain. In this sense, the TGU gain integral can be regarded as an optimization problem. We will consider nine degrees of freedom: electron betatron functions (2), X-ray Rayleigh ranges (2), electron transverse emittances (2), energy spread (1), frequency detuning (1), and TGU parameter (1). The rest of the machine parameters used in this study are given in Table~\ref{tab:machineparams}. These values are inspired by a typical 4GSR such as PETRA-IV~\cite{petra4, agapov2018}.

\begin{table}
  \caption{\label{tab:machineparams}Machine parameters used for optimization study. Storage ring parameters are derived from PETRA-IV~\cite{petra4, agapov2018}}
  \begin{ruledtabular}
  \begin{tabular}{l l l }
    Name & Symbol & Value \\
    \midrule
    \textbf{Storage ring} & & \\
    Electron energy & $E_{\text{beam}}$ & $5.96$ \si{GeV} \\
    Relativistic gamma & $\gamma_r$ & $1.167 \times 10^4$ \\
    Beam current (peak) & $I_{\text{pk}}$ & 31.89 \si{A} \\
    Relative energy spread & $\sigma_{\eta}$ & 0.1\% \\
    Natural emittance & $\varepsilon_{x,0}$ & 19 \si{pm\,rad} \\
    Emittance ratio & $k_c$ & 0.167 
                         \vspace{5pt}\\
    \textbf{Output radiation} & & \\
    %Res. wavelength & $\lambda_1$ & $0.861$ \si{\angstrom} \\
    Resonant energy & $\hbar \omega_1$ & $14.412$ \si{keV} \\
    Radiation emittance & $\varepsilon_r$ & $6.85$ \si{pm\,rad}
                                    \vspace{5pt} \\
    \textbf{Undulator} &  & \\
    %Undulator length & $L_u$ & 30 \si{m} \\
    Undulator period & $\lambda_u$ & 1.5 \si{cm} \\
    Number of periods & $N_u$ & 2000 \\
    Undulator parameter & $K_0$ & 1.06 \\
  \end{tabular}
  \end{ruledtabular}
\end{table}

\subsection{Optimization methodology}

We tested a number of different optimization algorithms on this problem, including gradient descent, simulated annealing and simple hill climber. We found that the objective function, i.e. the gain integral, had in all cases a clear global optimum and a simple convex shape. Provided reasonable starting parameters, all algorithms were able to converge relatively quickly (on the order of minutes) and reliably. The simple hill climber was chosen for its algorithmic simplicity.

\subsection{Optimization results}

\begin{figure}
  \includegraphics[width=\linewidth]{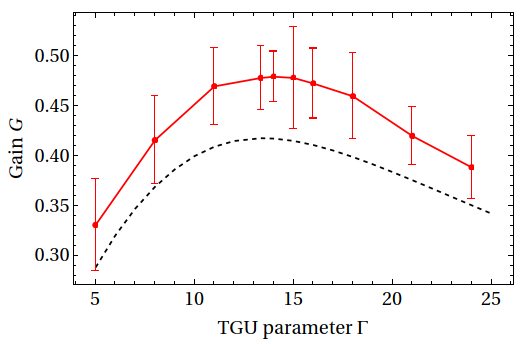}
  \caption{\label{fig:gvstgu1}Gain $G$ versus TGU parameter $\Gamma$ depicting results from numerical integration (black, dashed) and time-independent \texttt{GENESIS} simulation (red, solid). Error bars indicate 2 standard deviations of the shot-to-shot variation. Simulation results consistently overperform when compared to theory, likely due to the magnitude of gain $G\sim \mathcal{O}(1)$ surpassing the low-gain assumption of the analytical formula. Nevertheless, both models are consistent on the location of the gain optimum at $\Gamma = 13.3$.}
\end{figure}

The first parameter of interest is the TGU parameter $\Gamma$. Figure~\ref{fig:gvstgu1} shows TGU gain as a function of $\Gamma$, with the optimum value at $\Gamma_{\text{opt}} = 13.3$ and max gain $G_{\text{th}} = 0.42$ from numerical integration and $G_{\text{sim}} = 0.48$ from \texttt{GENESIS} simulation. The systematic overshoot from \texttt{GENESIS} simulation is a result of the magnitude of gain $G \sim \mathcal{O}(1)$ exceeding the low-gain assumption used to derive Eq.~\eqref{eq:gain}. In such cases, nonlinear interaction terms further boost its value in such a way that the gain increase $\propto G_{\text{th}}^2$ for $G<1$.  Nevertheless, the primary quantity of interest $\Gamma_{\text{opt}}$ remains consistent between simulation and theory.

\begin{figure}
  \includegraphics[width=\linewidth]{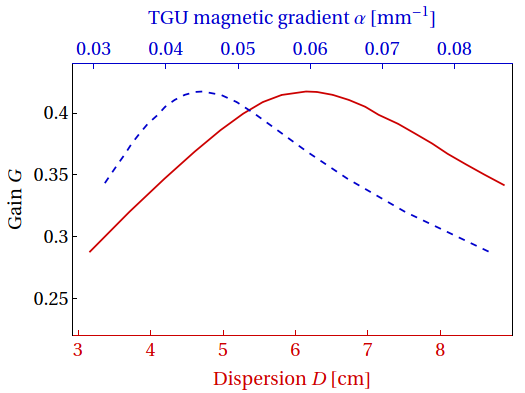}
  \caption{\label{fig:gvsdispamag1}Gain $G$ versus electron beam dispersion $D$ (red, solid) and TGU magnetic gradient $\alpha$ (blue, dashed) derived from numerical integration of the 3D gain formula. These parameters are derived from the TGU parameter $\Gamma$ by Eqs.~\eqref{eq:tgures2} and~\eqref{eq:tguparam}. Optimal gain is attained at $D_{\text{opt}} = 6.2\,\si{cm}$ and $\alpha_{\text{opt}} = 0.045 \, \si{mm^{-1}}$ for the chosen machine parameters.}
\end{figure}

We can convert $\Gamma_{\text{opt}}$ into the more practical values of dispersion $D$ and magnetic gradient $\alpha$ using equations~\eqref{eq:tgures2} and~\eqref{eq:tguparam}. Figure~\ref{fig:gvsdispamag1}  shows the optimization plots of these two parameters, with $D_{\text{opt}} = 6.2 \, \si{cm}$ and $\alpha_{\text{opt}} = 0.045 \, \si{mm^{-1}}$. The maxima are rather broad, affording flexibility to account for experimental limitations and/or imperfections.  For example, if we nominally define an ``acceptable'' gain value as falling within $10\%$ of the maximum, the allowable ranges of $D$ varies between $5 \text{ - } 7.5 \, \si{cm}$ and $\alpha$ between $0.035 \text{ - } 0.055 \, \si{mm^{-1}}$.

Note that although the optimal gradient $\alpha_{\text{opt}} = 0.045 \, \si{mm^{-1}}$ may be considered large, it still lays within the realm where the linear theory from Section II is appropriate. From the discussion around Eq.~\eqref{eq:tgures1}, we require that $\alpha y_j\ll 1$ for each individual electron, which in the ensemble sense translates to $\alpha \sigma_y \ll 1$, where $\sigma_y$ is the RMS electron beam size in $y$. This is amply satisfied in our case.

\begin{figure}
  \includegraphics[width=\linewidth]{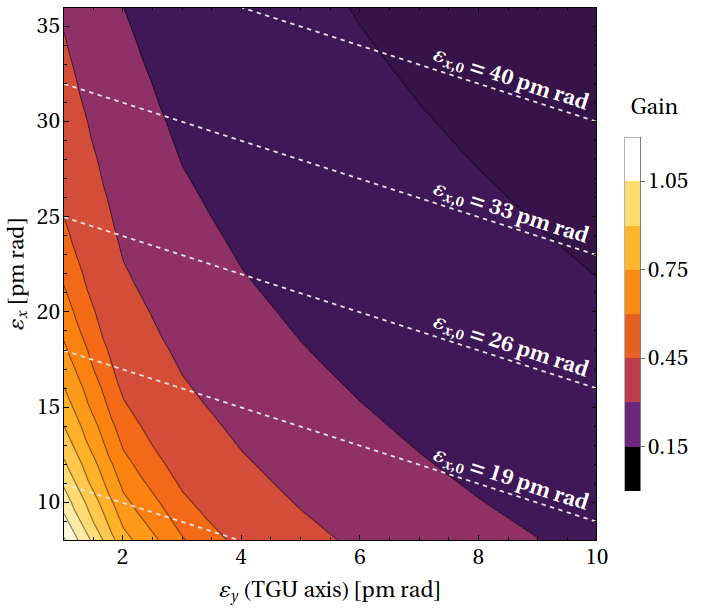}
  \caption{\label{fig:tguvsemittance}Contour plot of gain as a function of transverse emittances $\varepsilon_{x}, \varepsilon_y$. White dashed lines represent levels of constant natural emittance $\varepsilon_{x,0} \equiv \varepsilon_x + \varepsilon_y$. For fixed $\varepsilon_{x,0}$, reducing the emittance ratio $k_c \equiv \varepsilon_y/\varepsilon_x$ (hence reducing $\varepsilon_y$) results in significant gain improvement. Hence it would be advantageous for the TGU to minimize $k_c$ for given $\varepsilon_{x,0}$.}
\end{figure}

Next, we explore the effect of transverse emittances on TGU gain (Fig.~\ref{fig:tguvsemittance}). In a storage ring, the natural emittance $\varepsilon_{x,0}$ is determined by the magnetic lattice and radiation damping, with the individual emittances $\varepsilon_x, \varepsilon_y$ being set by tuning $k_c$. Unsurprisingly, we observe that TGU gain always increases as $\varepsilon_{x,0}$ is reduced, with the lower bound ultimately set by storage ring constraints. More importantly, however, the benefit of minimizing $\varepsilon_y$ far outweighs that of reducing $\varepsilon_x$. For instance, an increase of $1\, \text{pm rad}$ in $\varepsilon_y$ can have the same impact on TGU gain as an increase of $5$ to $10\,\text{pm rad}$ in $\varepsilon_x$. Thus, for a fixed $\varepsilon_{x,0}$, we should aim to minimize $k_c$ to maximally reap the benefits of the TGU.

\begin{figure*}
  \includegraphics[width=0.445\linewidth]{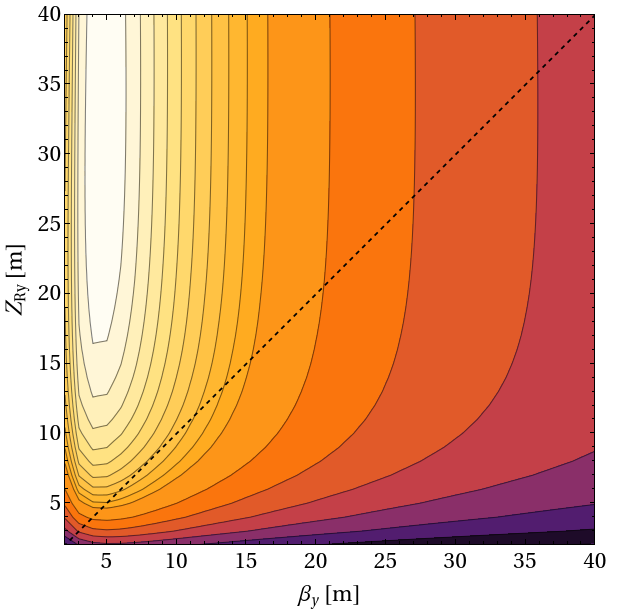}
  \includegraphics[width=0.49\linewidth]{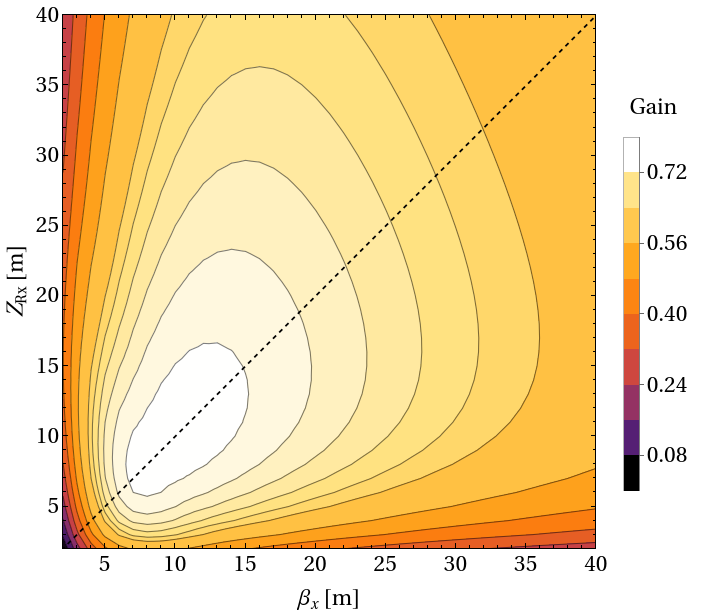}
  \caption{\label{fig:tguvsxy}Contour plot of gain vs electron and photon beam parameters in $x$ (right) and $y$ (left). In the $x$-direction, peak gain is located along the $\beta_x = Z_{Rx}$ line (black, dashed) as predicted by traditional FEL theory. In the $y$-direction, the contours are much more asymmetric due to the dispersion introduced by the TGU. In both directions, there are generous margins for potential tuning these beam parameters while still maintaining respectable gain.}  
\end{figure*}

Figure~\ref{fig:tguvsxy} shows the gain optimization plots with respect to the electron betatron function and X-ray Rayleigh length in the $x$ and $y$ dimensions respectively. In the $x$-axis, the optimal value lies along the $\beta_x = Z_{Rx}$ line. This agrees with planar FEL theory, which predicts that gain is maximized when the radiation mode size matches that of the electron beam~\cite{thebook}. On the other hand, the TGU gain contour is highly asymmetric in the $y$-dimension due to TGU dispersion. (To clarify, $\beta_y$ reflects the nominal betatron function before the introduction of dispersion.)

During storage ring operation, the beta functions $\beta_{x,y}$ can be constrained by lattice stability requirements. Using once again the $10\%$ gain dropoff threshold, we observe that $\beta_y$ and $\beta_x$ accept values up to $10\, \si{m}$ and $23\, \si{m}$ respectively. These ranges encompass the nominal figures for PETRA-IV insertion devices, with a generous margin for additional tuning if necessary.

The Rayleigh ranges $Z_{Rx}, Z_{Ry}$ are similarly constrained in reality, both by practical optical cavity design as well as user requirements. Fortunately, there is also a great deal of tunablility in these parameters. In the $x$-dimension, the optimal $Z_{Rx}$ is always equal to $\beta_x$, but a deviation of up to $\pm 10\, \si{m}$ still lays within the ``acceptable'' range. In the $y$-dimension, the allowance for sub-optimality is even larger, laying anywhere between $12\, \si{m}$ and $50 \, \si{m}$. Lastly, one special note should be made for the round-beam case ($Z_{Rx} = Z_{Ry}$), which may be desirable under certain circumstances. With this constraint imposed, we observed approximately 10\% dropoff from the optimal gain, making it feasible as an actual use case.

In summary, Table~\ref{tab:optparam1} lists some possible operation points for a TGU-enabled XFELO based on PETRA-IV parameters. Aside from the optimal TGU case, we include a weak dispersion scenario (when dispersion needs to be controlled), and a round-beam case (when the X-ray beam needs to have a symmetric profile). In all cases we find more than sufficient gain to drive a low-gain oscillator. 

\begin{table*}
  \caption{\label{tab:optparam1}Optimized TGU parameter sets. From left to right, the columns are gain from numerical integration $G_{\text{th}}$ and simulation $G_{\text{sim}}$, TGU parameter $\Gamma$, electron beta functions $\beta_y, \beta_x $, X-ray Rayleigh ranges $Z_{Ry}, Z_{Rx} $, detuning $x$, dispersion $D$, and TGU magnetic gradient $\alpha$.}
  \begin{ruledtabular}
  \begin{tabular}{l l l l l l l l l l l}
    Parameter set  & $G_{\text{th}}$ & $G_{\text{sim}}$ & $\Gamma$ & $\beta_y$ [m]  & $\beta_x$ [m] & $Z_{Ry}$ [m] & $Z_{Rx}$ [m] & $x$ & $D$ [cm] & $\alpha$ [$\si{mm^{-1}}$] \\
    \midrule
    Unconstrained TGU optimum  & 0.42 & 0.48 & 13.3 & 4.5 & 8.2 &  47.5 & 8.2 & 2.722 & 6.2 & 0.045 \\
    Low dispersion & 0.29 & 0.34 & 5 & 8.4 & 6.0 & 14.8 & 6.0 & 4.495 & 3.2 & 0.085 \\
    Round beam (constrain $Z_{Rx} = Z_{Ry}$) & 0.36 & 0.45 & 12 & 14.1 &  6.197 & 14.1 & 14.1 & 2.715 & 6.5 & 0.042 \\
  \end{tabular}
  \end{ruledtabular}
\end{table*}

\section{\label{sec:srxfelo}Implementation of a TGU-enabled storage ring XFELO}

While the previous section examined the TGU-enabled FEL in isolation, we would like now to focus on its implementation in a storage ring XFELO. Figure~\ref{fig:srxfelo} shows our proposed layout for the TGU-enabled storage ring XFELO. The design is made of three main components: (a) the storage ring, (b) the TGU, and (c) the X-ray cavity. The goal of this section is to demonstrate the performance of TGU-enabled SRXFELO using multi-stage numerical simulation, as well as discuss some implementation challenges regarding ring-FEL coupling. 

As before, we will base the machine parameters used in this study on proposed numbers for PETRA-IV~\cite{petra4, agapov2018}. While the PETRA-IV project does not include plans for a low-gain XFELO, we hope to demonstrate its feasibility for similar 4th generation facilities. Where appropriate, we have modified certain parameters to be more favorable to the XFELO. For instance, we reduced the emittance coupling factor and increased the peak electron current in order to boost TGU gain. A full summary of the machine parameters are provided in Tables~\ref{tab:machineparams} and~\ref{tab:simparams}.

\begin{table}
  \caption{\label{tab:simparams}Additional beam parameters used for simulation study. Unless otherwise stated, all other machine parameters are same as those provided in Table~\ref{tab:machineparams}.}
  \begin{ruledtabular}
  \begin{tabular}{l l l }
    Name & Symbol & Value \\
    \midrule
   \textbf{Electron} & & \\
    Storage ring circumference & $L_{\text{ring}}$ & 2333.87 \si{m} \\
    Emittance damping time ($y$) & $\tau_y$ & 22 \si{ms} \\
    Equilibrium emittance ($y$) & $\varepsilon_{y,0}$ & 2.714 \si{pm\,rad} \\ 
    Bunch charge (lasing) & $Q_b$ & 4 \si{nC} \\
    Bunch charge (non-lasing) & & 1 \si{nC} \\
    Number of lasing bunches & & 16 \\
    Betatron function at & & \\ \quad TGU midpoint ($x$) & $\beta_x$ & 8.2 \si{m} \\
    Betatron function at & & \\ \quad TGU midpoint ($y$) & $\beta_y$ & 4.5 \si{m} 
                             \vspace{5pt}\\
    \textbf{Cavity} & & \\
    Total length & $L_{\text{cavity}}$ & 145.87 \si{m}\\
    Round trip time & $T_{\text{cavity}}$ & 486.56 \si{ns}\\
    Focal length  & $f$ & 20.70 \si{m} \\
    Outcoupling at Crystal 1 & & 15\%  \\
    Rayleigh length at & & \\ \quad TGU midpoint ($x$) & $Z_{Rx}$ & 8.2 \si{m}\\
    Rayleigh length at & & \\ \quad TGU midpoint ($y$) & $Z_{Rx}$ & 8.2 \si{m}
                                    \vspace{5pt} \\
    \textbf{TGU} &  & \\
    Dispersion & $D$ & 5.05 \si{cm} \\
    Magnetic gradient & $\alpha$ & 0.055 \si{mm}$^{-1}$ \\
  \end{tabular}
  \end{ruledtabular}
\end{table}

The TGU parameters are informed by our optimization results in the previous section. Some modifications have been made, most notably to the X-ray Rayleigh ranges to accomodate the X-ray cavity. The cavity is laid out in the bow-tie configuration, derived from~\cite{cavity}. We chose the stable mode to be a symmetric Gaussian mode with $Z_R = 8.2 \, \si{m}$ measured from the midpoint of the TGU. This ensures that the cavity would have realistic dimensions, and the requisite focal lengths of the lenses are feasible using off-the-shelf beryllium compound refractive lenses (CRLs)~\cite{crlpaper}. 

\subsection{Simulation methodology}

\begin{figure}
   \includegraphics[width=0.9\linewidth]{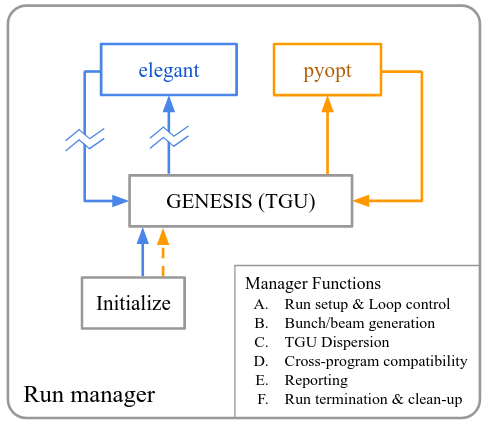}
  \caption{\label{fig:simschematic}Flowchart of simulation pipeline used in study. Blue lines indicate electron simulation pathway, while orange lines indicate X-ray pathway. Dashed line indicates optional step. The run manager serves as a wrapper program that provides functions for initialization, beam generation/manipulation, reporting, and clean-up, amongst other things. The main FEL simulation is handled by modified \texttt{GENESIS} version 2 (see main text for details of modification), while \texttt{elegant} handles the storage ring propagation. Broken lines between $\tt GENESIS$ and $\tt elegant$ indicate that the electron file is not preserved between iterations (see text for discussion). X-ray cavity simulation is handled by an in-house Fourier-based code named $\tt pyopt$. The X-ray file is preserved throughout. }
\end{figure}

A custom simulation framework was written for this study (Fig.~\ref{fig:simschematic}). The simulation pipeline can be broken down into four parts, namely (a) overall run management, (b) TGU-FEL simulation, (c) storage ring propagation, and (d) X-ray cavity propagation .

Run management is handled by a wrapper program. The user specifies the overall run parameters via a parameter input file. The run manager interprets the input file and initializes the run accordingly, including generating the specified electron bunch and photon beam files (latter optional). During each run loop, the run manager ensures consistency with the input parameter files of $\tt elegant$/$\tt GENESIS$, followed by calling the respective program at the appropriate step. Between loops, the manager performs the requisite minor beam/bunch manipulations, such as TGU dispersion and bunch re-generation (more discussed later). The run manager also collects and reports on beam/bunch statistics. At the end of the run, the manager is responsible for termination and clean-up.

The TGU-FEL simulation is handled by modified $\tt GENESIS$ (version 2 in $\tt Fortran$), capable of both single-slice and 3D time-dependent run modes. The requisite TGU modification is made in the $\tt faw2$ method located in source file $\tt magfield.f$. The $\tt faw2$ method is responsible for calculating the squared off-axis value of the undulator parameter $a_u^2 \equiv K^2/2$~\cite{genesismanual}. We modify as follows:

\begin{verbatim}
  faw2 = awz(i)*awz(i)*(1.d0
           + 2.d0*atgu*xt+ atgu*atgu*xt*xt
           + xkx*xt*xt + xky*yt*yt)
\end{verbatim}

The middle line of code is our addition. We introduce the variable $\tt atgu$ $\equiv \alpha/k_u$ representing the linear gradient of the TGU as per Eq.~\eqref{eq:tgugradientlinear}. (The additional factor of $1/k_u$ comes from coordinate normalization in $\tt GENESIS$.) Note that the TGU gradient is applied in the $x$-direction because of the built-in assumption in $\tt GENESIS$ that the undulator is aligned horizontally. To compensate, we will need to rotate the electron and X-ray distributions using the run manager before and after each $\tt GENESIS$ run. At the same time, we will also introduce the requisite dispersion in the electron bunch.

Storage ring propagation is handled by $\tt elegant$. Due to the discrete slice representation of the electron bunch in $\tt GENESIS$, the particle file is not easily made compatible with $\tt elegant$. Instead, we track the first- and second-order moments of the bunch after each $\tt GENESIS$ run, and rely on $\tt elegant$'s built-in generator $\tt bunched\_beam$ to re-generate the bunch accordingly~\cite{elegantmanual}. The electrons are then propagated using a one-turn lattice file with periodic transport ($\tt ILMATRIX$ element), synchrotron radiation damping ($\tt SREFFECTS$), and RF acceleration ($\tt RFCA$). After that, the bunch moments are recorded and a $\tt GENESIS$ bunch is generated by the run manager accordingly.

X-ray cavity propagation is handled by an in-house Fourier-based code, named $\tt pyopt$. The code accounts for basic effects such as focusing and diffraction, as well as full frequency-angle bandwidth filtering of the diamond Bragg reflectors, along with (optional) phase errors and mirror misalignments. The cavity code uses an identical representation of the X-ray field file as $\tt GENESIS$ so that the field can be tracked throughout the entire run.

One final note about the run loop control: the X-ray cavity length is typically a fraction of the storage ring circumference. In our study, this ratio is $1/16$, i.e. the photon pulse traverses the cavity 16 times for each orbit of the electron bunch. Alternatively, the photon pulse interacts with 16 consecutive electron bunches evenly spaced around the ring. The latter scheme is chosen for this study. At any given time, the run manager keeps track of the beam moments of all 16 bunches seperately and selects the appropriate bunch prior to each $\tt GENESIS$ run. 

\subsection{Simulation results}

\begin{figure}
   \includegraphics[width=\linewidth]{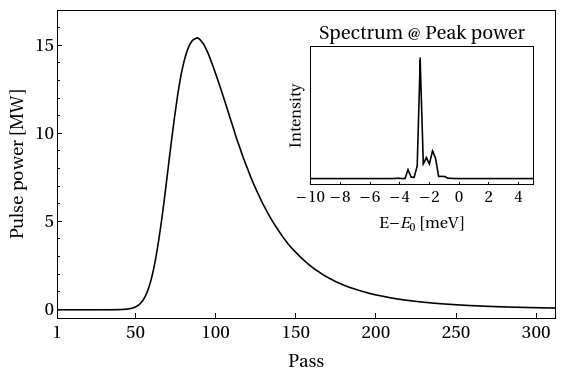}
  \caption{\label{fig:pulsepower}Plot of XFELO pulse power vs. turn number from numerical simulation. Maximum power of $15 \, \si{MW}$ is achieved at turn 95. The inset shows the frequency spectrum at peak power. The spectrum has a full-width half-maximum (FWHM) of $2 \, \si{meV}$ corresponding to a relative bandwidth of $1.4 \times 10^{-7}$.}
\end{figure}

We report on the simulation results using the framework discussed previously and the parameters of Table~\ref{tab:simparams}. Figure~\ref{fig:pulsepower} shows the power evolution of one XFELO pulse. Peak intracavity power of $15 \, \si{MW}$ was achieved at turn $95$ after initial onset. With $15\%$ thin-mirror outcoupling, this corresponds to $2.25 \, \si{MW}$ of usable power. At peak power, the photon bandwidth has a full-width half-maximum (FWHM) of approximately $2 \, \si{meV}$ (relative bandwidth of $1.4 \times 10^{-7}$).

\begin{figure}
  \includegraphics[width=0.8\linewidth]{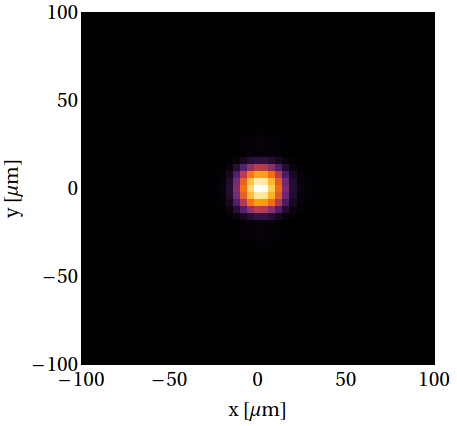}
  \caption{\label{fig:beamspot}X-ray beamspot at peak power. The root-mean-square (RMS) beam sizes are $\sigma_x = 14 \, \si{\mu m}$ and $\sigma_y = 12 \, \si{\mu m}$. The slight asymmetry, despite the symmetric cavity, is due to the Bragg crystal filtering acting only in the horizontal plane.}
\end{figure}

Contrary to previous studies~\cite{oscillatorprev}, we observed multiple peaks within the overall photon bandwidth, rather than a single one. We attribute this difference to the length of the electron bunch (or simulation window in the case of constant bunch current). In previous studies, a relatively short bunch with $\sigma_t \lesssim 2 \, \si{ps}$ was used. This is, however, not feasible under current circumstances, where IBS and Touschek lifetime considerations limit realistic values of root-mean-square (RMS) bunch length to $20$ to $50 \, \si{ps}$ for the desired peak currents. 

In order to conserve simulation time and memory, we chose a $20\, \si{ps}$ simulation window with a constant current profile. On average, we observed $ \sim 4$ peaks within the FWHM bandwidth. For longer electron bunch lengths, we predict that more peaks would arise within the same overall bandwidth, with the width of each peak set by the Fourier limit. In order to verify this, we ran a number of additional numerical experiments of low-gain oscillators, both with and without the TGU. We found empirically that the total FWHM bandwidth in saturation has a lower bound between five to ten times smaller than the crystal bandwidth $\sigma_\omega$. For electron bunch lengths shorter than $\sim 5/\sigma_\omega$, the output has one single spectral peak as a result of the Fourier limit. On the other hand, longer bunches can result in multiple spectral modes within the overall bandwidth of $\sigma_\omega/5$. Currently, we do not have a rigorous theory to indicate if this limit is fundamental or to what extent it may be changed; we plan to do further studies of this phenomenon in the future.

In the bowtie cavity, we observed maximum TGU gain of approximately $G_{\text{TGU}} = 0.3$, after accounting cavity losses and intentional outcoupling. This includes losses at all four Bragg crystals. Additional losses at other optical components, such as CRLs, as well as losses due to misalignments and crystal phase front errors were not taken into account in this study. However, the net round trip gain of $0.3$ provides a healthy margin for these additional losses.

Figure~\ref{fig:beamspot} shows the beam profile at peak power, with RMS beam sizes of $\sigma_x = 14 \, \si{\mu m}$ and $\sigma_y = 12 \, \si{\mu m}$. While the nominal beam is intended to be round ($\sigma_x = \sigma_y$), the slight asymmmetry is attributed to the angular filtering of the Bragg crystal, which only takes place in the horizontal plane of the cavity. The narrow angular bandwidth imposed by the crystals results in a corresponding increase in the beam size in $x$.

\subsection{Ring-FEL coupling and other implementation challenges}

The X-ray cavity and storage ring constitute a coupled oscillator system. The physics of the storage ring FEL has been previously studied in the UV/IR regimes~\cite{srxfelotheory, srxfelochaos, srxfelotheory2}. Though the energies involved here are greater, much of the theory, especially relating to the ring-FEL interaction, remains relevant in the X-ray regime. For the sake of the following discussion, we provide here a short summary of the most salient points adapted from~\cite{srxfelotheory, srxfelotheory2}.

In a storage ring FEL, there exists a tension between the electrons circulating in the ring and the photons circulating in the cavity. Namely, as photon intensity increases, the electron bunch suffers increasing degradation of its energy spread. This in turn leads to the reduction of FEL gain, and eventually, the saturation and decline of laser power. The addition of the TGU modifies this interaction in one important aspect --- electron beam degradation takes place in the TGU-axis transverse space ($y$ in our case), rather than longitudinally. This is because of the dispersive sections before and after the undulator, which effectively translates the energy spread growth generated by the FEL into excess emittance in the ring.

Hence, the macrotemporal dynamics of the TGU-enabled storage ring FEL can be described by the following linearized equations:
\begin{eqnarray}
  \frac{dU}{dt} &=& U \frac{g-L}{\theta} + U_s, \label{eq:macro1} \\
  \frac{d(\varepsilon_y)}{dt} &=& -\frac{2}{\tau_y}(\varepsilon_y - \varepsilon_{y,0}) + \Delta_U, \label{eq:macro2} \\
  g &=& g_0 \exp(-k(\varepsilon_y - \varepsilon_{y,0}))\left[ 1+ F(t) \right]. \label{eq:macro3}
\end{eqnarray}
Here, $U$ is the FEL intensity, $g,L,\theta$ are the gain, loss and transit time for each turn respectively, $U_s$ is the intensity of the spontaneous emission, $\varepsilon_{y,0}$ is the equilibrium $y$ emittance, $\tau_y$ is the characteristic emittance damping time, $\Delta_U$ is the increase in emittance due to laser intensity, $g_0$ is the maximum gain at equilibrium emittance, and $F(t)$ represents an optional external gain modulation imposed on the system.

\begin{figure}
   \includegraphics[width=\linewidth]{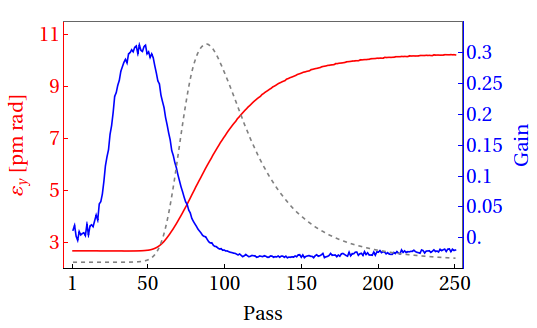}
  \caption{\label{fig:beamdegradation}Plot depicts the degradation of electron emittance $\varepsilon_y$ due to FEL gain within a single oscillator pulse. Initially, the low emittance $\varepsilon_y$ (red) enables a large positive gain (blue), which in turn exponentially amplifies the X-ray power (dashed gray; vertical scale not shown). The increasing X-ray power degrades $\varepsilon_y$ due to the TGU-FEL interaction. Near saturation, emittance $\varepsilon_y$ has increased so much that gain falls below zero, leading to the exponential decay of the X-ray pulse.}
\end{figure}

Figure~\ref{fig:beamdegradation} depicts the evolution of these three parameters over a single FEL pulse. At first, the photon intensity is dominated by spontaneous emission, which quickly gets overtaken by FEL amplification as the equilibrium emittance $\varepsilon_{y,0}$ enables the maximum theoretical TGU gain $g_0$. The exponential rise in photon intensity results in increasing degradation of $\varepsilon_y$, which causes $g$ to suffer. Finally, at saturation, $g$ falls below single turn loss $L$ (such that $g-L \leq 0$), which marks the start of the exponential decay of the photon pulse. This whole process takes place over hundreds of oscillator turns (typically $\sim 10^{-4}\, \si{s}$).

On the much longer timescale, the electron bunch damps in $\varepsilon_y$ as it circulates within the storage ring (typical values of $\tau_y \sim 10^{-2} \, \si{s}$). This damping process is well understood in the realm of storage ring physics~\cite{msands, syphers}. In short, the electrons undergo spontaneous radiation emission as they traverse the bending sections (and radiation insertion devices) of the circular ring. Each photon emission leads to a slight loss in momentum $\delta \va*{p}$ approximately parallel to the instantaneous momentum vector $\va*{p}$ of the electron. In general, this is not parallel to the nominal orbit trajectory, as each electron undergoes stable betatron oscillation around the nominal axis.

Every turn, the electron passes through RF cavities which ``top-up'' the lost energy due to spontaneous emission. The momentum gain from the RF cavities \emph{is} parallel to the nominal axis. This means that, on balance, the electron would lose some of its transverse momentum $\va*{p}_{\perp}$ every turn. On the ensemble level, this translates to the gradual shrinking of the transverse emittances of an electron bunch.

This damping does not continue indefinitely, however. Due to its quantum nature, each packet of spontaneous emission will excite random betatron oscillations in the electron. These random oscillations, when propagated through the magnetic lattice of the storage ring, are responsible for the equilibrium emittances of the stored bunch. This effect is essentially constrained to the bending plane, so that quantum excitation sets the emittance along the $x$-axis.  On the other hand, the emittance in $y$ is typically able to reach much smaller values (by up to several orders of magnitude), with its equilibrium value ultimately being decided by (intentional or not) coupling between the two transverse axes.

Thus, the ring-FEL system behaves like a coupled oscillator with two drastically different timescales. On the one hand, the laser is characterized by its rise time $\tau_0 \sim 10^{-6} \, \si{s}$. On the other hand, emittance damping is set by $\tau_y \sim 10^{-2}\, \si{s}$. Using Eqs.~\eqref{eq:macro1} --~\eqref{eq:macro3}, we can determine the natural period of the system to be 
\begin{equation}
  T_R = 2\pi \sqrt{\frac{\tau_0\tau_y}{2}}.
\end{equation}
The system is extremely sensitive to initial conditions, displaying chaotic and noisy behaviour close to this natural resonance~\cite{dukefel, srxfelochaos}. A more stable alternative can be achieved by modulating the gain periodically in the FEL so as to effectively ``turn on'' and then ``turn off'' the XFELO. Thus, the laser can operate in a pulsed fashion with greater stability. The duty cycle of the modulation is given by: (a) ``off'' state for several $\tau_y$ damping times (tens to hundreds of milliseconds), then (b) ``on'' state for several $\tau_0$ laser rise times (tens of microseconds) until saturation. The laser switches itself off due to emittance degradation at saturation.

Gain modulation can be achieved by essentially disrupting the temporal and/or spatial overlap between the electron bunch and the photon pulse, thus suppressing the FEL interaction. For our proposed parameters, a temporal displacement of $\sim 20$ to $50 \, \si{fs}$ between consecutive bunches is sufficient to ``switch off'' the XFELO. This can be done by detuning the storage ring RF system to the order of tens of Hz. This method has been experimentally tested, albeit on a smaller scale, at the Duke storage ring FEL \cite{dukefel}. Other potential gain modulation methods include transversely displacing the electron trajectory in the undulator (e.g.~with a kicker) or manipulating the X-ray cavity length/geometry, although to perform the latter at the required millisecond time scale without compromising cavity stability remains an open area of research.

One additional note must be made about the ring-FEL coupling strategy. Since the storage ring circumference is typically many times the length of the X-ray cavity, it is necessary to have multiple electron bunches spaced around the ring with temporal separation matching the cavity round-trip time. In the numerical study, we had a ring-cavity circumference ratio of 16, meaning that we had 16 of these ``XFELO bunches''. This is a fraction of the hundreds, if not thousands, of simultaneously circulating bunches present in a modern storage ring.

As discussed previously, the XFELO lasing process leads to significant emittance degradation due to the elevated laser intensity. We would like to isolate this detrimental effect to only the XFELO bunches while leaving the non-XFELO bunches relatively unperturbed. The problem of selecting for the XFELO bunches is non-trivial~\cite{tgu2013}.

One commonly proposed technique is to situate the XFELO on a bypass with fast transverse kickers, which periodically kick the XFELO bunches into the oscillator. This is not feasible when the closely spaced electron bunches are only separated by gaps on the order of single to tens of nanoseconds. Even if the rise/fall time of the kicker fits within the bunch separation, its repetition rate is not fast enough to support the oscillator~\cite{kicker}.

\begin{figure}
   \includegraphics[width=\linewidth]{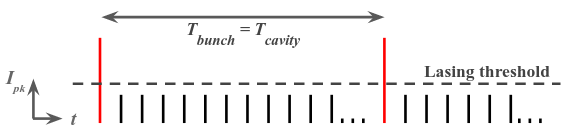}
  \caption{\label{fig:macrotemporalstructure}Proposed bunch train structure for ring-FEL coupling. Each vertical line represents one bunch, with its height proportional to peak current $I_{\text{pk}}$ and thus TGU gain by Eq.~\eqref{eq:gainprefactor}. Tall red bunches are the designated ``XFELO bunches'' with $I_{\text{pk}}$ exceeding the lasing threshold set by single-turn cavity loss. The temporal gap between XFELO bunches $T_{\text{bunch}}$ is equal to the X-ray round-trip time $T_{\text{cavity}}$. The non-XFELO bunches (black) lay below the lasing threshold and only experience spontaneous undulator radiation. Thus they are able to avoid substantial emittance degradation from the XFELO lasing process.}
\end{figure}

Our current proposal is to utilize bunch charge stacking (Fig.~\ref{fig:macrotemporalstructure}). Under this scheme, the XFELO bunches, spaced at the appropriate periodic interval, are intentionally injected with multiple times the nominal bunch charge, while adjacent RF buckets are left empty to compensate for the beam loading effect. This technique has precedence in the APS-U, where it is employed to mitigate ion instabilities~\cite{apsu}.

Since TGU gain is linearly proportional to peak current in the low gain approximation, the XFELO bunches will experience proportionally more gain than the non-XFELO ones. Then, it suffices to set the single turn loss within the cavity (for instance, using outcoupling methods) to be higher than the gain experienced by non-XFELO bunches, but lower than that of the XFELO bunches. Thus only the latter will experience exponential growth and the subsequent emittance degradation due to the intense laser field near saturation. The non-XFELO bunches will only undergo spontaneous undulator radiation, which will not cause substantial emittance increase.

In our numerical study, we chose the non-XFELO bunches to have a total charge of $Q_b = 1 \, \si{nC}$, with a corresponding peak current of $I_{\text{pk}} = 10 \, \si{A}$, whereas the stacked XFELO bunches have $Q_b = 4 \, \si{nC}$ and $I_{\text{pk}} = 32 \, \si{A}$, after accounting for bunch lengthening due to intra-beam scattering (IBS) and ring impedance. Single turn loss is set to approximately $0.15$ via thin mirror outcoupling.

\section{\label{sec:conclusion}Conclusion}
In this study, we have explored the feasibility of a low-gain, TGU-enabled XFELO driven by a modern 4th generation electron synchrotron. We derived the TGU gain formula in the low gain limit, and then performed gain optimization of several key beam parameters. Finally, we selected a near-optimal parameter set to study under multi-stage simulation, in order to obtain projected performance figures.

The results demonstrate that the TGU-XFELO, under pulsed-mode operation, can be a feasible inclusion for future storage rings. Moreover, the parameter set chosen for this study is just one of many possible configurations. The mathematical and numerical framework developed for this study provides a useful toolkit for modeling future TGU-XFELO design under a wide range of practical circumstances.

\begin{acknowledgments}
   The numerical simulations in this study make use of codes written and maintained by S. Reiche (\texttt{GENESIS 1.3}) and M. Borland (\texttt{elegant}). The authors would also like to thank G. Tiwari and J.-W. Park for advice and fruitful discussions. This work is supported by the U.S. Department of Energy, Office of Science under Contract No. DE-AC02-06CH11357.
\end{acknowledgments}
\vspace{0.25in}

\appendix

\section{Derivation of 3D TGU Gain Formula}

We start from Eq.~\eqref{eq:convformula}.

\begin{widetext}
For future convenience, we perform a change of variables $\va*{\phi}-\va*{p} \rightarrow \va*{\phi}$ and $\va*{x}-\va*{y} \rightarrow \va*{x}$. We also use integration by parts to switch the derivative, thus obtaining
\begin{eqnarray}
  G &=& -\frac{G_0}{8\pi N_uL_u^2\lambda_1^2} \int d\eta d \va*{x} d \va*{y} d\va{\phi} d\va*{p} \, B_E(\va*{y}, \va*{\phi} + \va*{p})\pdv{\eta} B_U(\eta, \va*{x}, \va*{\phi}, \va*{p}, \va*{y}) F(\eta, \va*{x} + \va*{y}, \va*{p}).
\end{eqnarray}
Since the two transverse dimensions are decoupled, we drop the vector notation and focus on the TGU dimension $y$. (The $x$ case can be easily obtained later by setting $\Gamma \rightarrow 0$). At this point, note that $x,y$ will refer to spatial integration variables instead of the usual transverse dimensions. The Gaussian X-ray seed is given by
\begin{equation}
  B_E(y, \phi+p) = \frac{1}{ 2\pi \sigma_{r}\sigma_{\phi}} \exp \left[ - \frac{y^2}{2\sigma_{r}^2} - \frac{(\phi+p)^2}{2\sigma_{\phi}^2} \right].
  \label{eqapp:be}
\end{equation}
Here, $\sigma_r, \sigma_{\phi}$ refers to the RMS X-ray beam size and divergence in $y$ respectively. The electron distribution is given by
\begin{equation}
  F(\eta, x+y, p) = \frac{1}{(2\pi)^{3/2}\sigma_y\sigma_{\eta}\sigma_{p} } \exp \left[ - \frac{(x+y-D\eta)^2}{2\sigma_y^2} - \frac{\eta^2}{2\sigma_{\eta}^2} - \frac{p^2}{2\sigma_{p}^2} \right].
  \label{eqapp:f}
\end{equation}
Here, $\sigma_y,\sigma_p$ refers to the RMS electron beam size and divergence in $y$ respectively, and $\sigma_{\eta}$ refers to the normalized energy spread where $\eta \equiv (\gamma-\gamma_0)/\gamma_0$. As discussed in the main text, the spontaneous undulator brightness can be obtained from the Wigner transform of Eq.~\eqref{eq:brightnessundulator}:
\begin{eqnarray}
  B_U(\eta,x,\phi,p,y) &=& \int^{L/2}_{-L/2} dzds\, \exp \left[ ik_u\Delta\nu(z-s)- 2ik_u\eta(z-s)   + i k_u T_{\alpha} \left\{ (x+y)(z-s) + \frac{p}{2}(z^2-s^2) \right\}  \right] \nonumber \\ && \quad \times \int d\xi\, \exp \left[ -ikx\xi + \frac{ik}{2} \Big( (\phi + \xi/2)^2z - (\phi - \xi/2)^2s \Big) \right],
\end{eqnarray}
where $\Delta \nu \equiv (\omega-\omega_1)/\omega_1 $ is the frequency detuning from the fundamental harmonic. We evaluate the Wigner integral to get
\begin{eqnarray}
 B_U(\eta,x,\phi,p,y) &=& \int^{L/2}_{-L/2} dzds\, \sqrt{\frac{8\pi i}{k(z-s)}} \exp \left[ ik_u\Delta\nu(z-s)- 2ik_u\eta(z-s) \right. \nonumber \\ && \quad \left. + i k_u T_{\alpha} \left\{ (x+y)(z-s) + \frac{p}{2}(z^2-s^2) \right\} - \frac{2ik}{z-s} x^2 - \frac{2iksz}{z-s}\phi^2 + \frac{2ik(z+s)}{z-s}x\phi  \right]. 
\end{eqnarray}

\end{widetext}
Note that
\begin{equation}
  \pdv{B_U}{\eta} = -2ik_u(z-s)B_U.
  \label{eqapp:dbu}
\end{equation}
We then substitute Eqs.~\eqref{eqapp:be} through \eqref{eqapp:dbu} into the gain convolution formula, and perform Gaussian integration in each variable. Generically, the Gaussian integral takes the form
\begin{equation}
  \int dx \exp(-Ax^2 + Bx) = \sqrt{\frac{\pi}{A}} \exp \left( \frac{B^2}{4A} \right).
  \label{eqapp:gaussian}
\end{equation}
In each case, we will use coefficients $A, B$ with the appropriate subscripts (e.g. $A_x, B_x$ for the $dx$ integral) to denote the result. For brevity, we will define the following parameters:
\begin{eqnarray}
  \Sigma_y^2 &=& \sigma_y^2 + \sigma_r^2 + D^2\sigma_{\eta}^2, \\
  \Sigma_{yr}^2 &=& \sigma_{y}^2 + \sigma_{r}^2, \\
  \Sigma_{y\eta}^2 &=& \sigma_{y}^2 + D^2\sigma_{\eta}^2, \\
  \Sigma_{\phi}^2 &=& \sigma_p^2 + \sigma_{\phi}^2.
\end{eqnarray}
We will also use short-hand notation for the following recurring terms:
\begin{eqnarray}
  \left[\hdots \sigma_{\phi}^2\right] &=& (z-s) + 4ik sz \sigma_{\phi}^2, \\
  \left[\hdots \Sigma_{\phi}^2 \right] &=& (z-s) + 4ik sz \Sigma_{\phi}^2, \\
  \left[BD\right] &=& (z-s)[1 + 4k^2 \Sigma_{yr}^2\Sigma_{\phi}^2] \nonumber \\ && \qquad + 4ik [\Sigma_{yr}^2 + sz \Sigma_{\phi}^2], \\
  \left[BD \right]_y &=& (z-s)[1 + 4k^2 \Sigma_{y}^2\Sigma_{\phi}^2] \nonumber \\ && \qquad + 4ik [\Sigma_{y}^2 + sz \Sigma_{\phi}^2], \\
  \left[BD\right]_{\sigma_r} &=& (z-s)[1 + 4k^2 \sigma_{r}^2\Sigma_{\phi}^2] \nonumber \\ && \qquad + 4ik [\sigma_{r}^2 + sz \Sigma_{\phi}^2], \\
  \left[BD\right]_{\sigma_{\phi}} &=& (z-s)[1 + 4k^2 \Sigma_{y}^2\sigma_{\phi}^2] \nonumber \\ && \qquad + 4ik [\Sigma_{y}^2 + sz \sigma_{\phi}^2].
\end{eqnarray}
We will perform a total of five Gaussian integrals, followed by consolidating and simplifying the prefactor and the terms in the exponential. The steps are listed in order below.

\subsection{The $d\phi$ integral}
The integral takes the form
\begin{eqnarray}
   \int d\phi \exp \left[ - \phi^2 \left( \frac{1}{2\sigma_{\phi}^2} + \frac{2iksz}{z-s} \right) \right. && \nonumber  \\  \quad \left. + \phi \left( - \frac{p}{\sigma_{\phi}^2} + \frac{2ik(z+s)x}{z-s} \right) \right], &&
\end{eqnarray}
whence we can consolidate the terms
\begin{eqnarray}
  A_{\phi} &=& \frac{1}{2\sigma_{\phi}^2} + \frac{2iksz}{z-s} = \frac{[\hdots \sigma_{\phi}^2]}{2(z-s)\sigma_{\phi}^2}, \\
  B_{\phi} &=& \frac{2ik(z+s)x\sigma_{\phi}^2 - p(z-s)}{\sigma_{\phi}^2(z-s)}, \\
  \frac{B_{\phi}^2}{4A_{\phi}} &=& p^2\, \frac{z-s}{2\sigma_{\phi}^2[\hdots \sigma_{\phi}^2]} - px \,\frac{2ik(z+s)}{[\hdots \sigma_{\phi}^2]} \nonumber \\ && \qquad - x^2\, \frac{2k^2(z+s)^2\sigma_{\phi}^2}{(z-s)[\hdots \sigma_{\phi}^2]}.
\end{eqnarray}
From the form of Eq.~\eqref{eqapp:gaussian}, we note that $A_{\phi}$ will feature in the prefactor of the final result, whereas the $B^2_{\phi}/4A_{\phi}$ term will carry into subsequent integrations. This is repeated for the subsequent integration steps.

\subsection{The $dy$ integral}
We have
\begin{eqnarray}
  &&\int dy \exp \left[ -y^2 \left( \frac{1}{2\sigma_r^2} + \frac{1}{2\sigma_y^2} \right) \right. \nonumber \\ && \quad \left. + y \left( - \frac{x-D\eta}{\sigma_y^2} + ik_uT_{\alpha}(z-s) \right)\right],
\end{eqnarray}

\begin{widetext}
whence
\begin{eqnarray}
  A_y &=& \frac{\sigma_y^2 + \sigma_r^2}{2\sigma_y^2\sigma_r^2} = \frac{\Sigma_{yr}^2}{2\sigma_y^2 \sigma_r^2}, \\
  B_y &=& \frac{ik_uT_{\alpha}(z-s)\sigma_y^2 - x + D\eta}{\sigma_y^2}, \\
  \frac{B_y^2}{4A_y} &=& x^2\, \frac{\sigma_r^2}{2\sigma_y^2\Sigma_{yr}^2} - x\eta\, \frac{D\sigma_r^2}{\sigma_y^2\Sigma_{yr}^2} + \eta^2 \, \frac{D^2\sigma_r^2}{2\sigma_y^2\Sigma_{yr}^2} - x\, \frac{ik_uT_{\alpha}(z-s)\sigma_r^2}{\Sigma_{yr}^2 } \nonumber \\ && + \eta \, \frac{iDk_uT_{\alpha}(z-s)\sigma_r^2}{\Sigma_{yr}^2} - \frac{\left[ k_uT_{\alpha}(z-s)\sigma_r\sigma_y \right]^2}{2\Sigma_{yr}^2}.
\end{eqnarray} 

\subsection{The $dp$ integral}

Including terms from the $d\phi$ integral, we get
\begin{equation}
   \int dp  \exp \left[ -p^2 \left( \frac{1}{2\sigma_{\phi}^2} + \frac{1}{2\sigma_p^2} - \frac{z-s}{2\sigma_{\phi}^2[\hdots \sigma_{\phi}^2]} \right)  + p \left( \frac{ik_uT_{\alpha}(z^2-s^2)}{2} - \frac{2ik(z+s)x}{[\hdots \sigma_{\phi}^2]} \right) \right],
\end{equation}
whence
\begin{eqnarray}
  A_p &=& \frac{(z-s) + 4iksz \Sigma_{\phi}^2}{2\sigma_p^2[\hdots \sigma_{\phi}^2]}= \frac{[\hdots \Sigma_{\phi}^2]}{2\sigma_p^2[\hdots \sigma_{\phi}^2]}, \\
  B_p &=& \frac{ik_uT_{\alpha}(z^2 - s^2) [\hdots \sigma_{\phi}^2] - 4ik(z+s)x}{2[\hdots \sigma_{\phi}^2]}, \\
  \frac{B_p^2}{4A_p} &=& -x^2\, \frac{2k^2(z+s)\sigma_p^2}{[\hdots \Sigma_{\phi}^2][\hdots \sigma_{\phi}^2]} + x \, \frac{kk_uT_{\alpha}(z+s)(z^2-s^2)\sigma_p^2}{[\hdots \Sigma_{\phi}^2]} - \frac{[k_uT_{\alpha}\sigma_p(z^2-s^2)]^2[\hdots \sigma_{\phi}^2]}{8[\hdots \Sigma_{\phi}^2]}.
\end{eqnarray} 

\subsection{The $dx$ integral}

The previous three integrations all contribute terms to this integral:
\begin{eqnarray}
  && \int dx \exp \left[ -x^2 \left( \frac{1}{2\sigma_y^2} + \frac{2ik}{z-s}  + \frac{2k^2(z+s)^2\sigma_{\phi}^2}{(z-s)[\hdots \sigma_{\phi}^2]}  - \frac{\sigma_r^2}{2\sigma_y^2\Sigma_{yr}^2} + \frac{2k^2(z+s)^2\sigma_p^2}{[\hdots\Sigma_{\phi}^2][\hdots \sigma_{\phi}^2]} \right) \right. \nonumber \\ && \qquad \left. + x \left( \frac{D\eta}{\sigma_y^2} + ik_uT_{\alpha}(z-s)- \frac{D\eta\sigma_r^2}{\sigma_y^2\Sigma_{yr}^2} - \frac{ik_uT_{\alpha}(z-s)\sigma_r^2}{\Sigma_{yr}^2} + \frac{kk_uT_{\alpha}(z+s)(z^2 - s^2)\sigma_p^2}{[\hdots \Sigma_{\phi}^2]} \right) \right],
\end{eqnarray}
whence
\begin{eqnarray}
  A_x &=& \frac{(z-s)[1 + 4k^2\Sigma_{yr}^2\Sigma_{\phi}^2] + 4ik[\Sigma_{yr}^2 + sz\Sigma_{\phi}^2]}{2\Sigma_{yr}^2[\hdots \Sigma_{\phi}^2]} = \frac{[BD]}{2\Sigma_{yr}^2[\hdots \Sigma_{\phi}^2]}, \\
  B_x &=& \frac{(ik_uT_{\alpha}(z-s)\sigma_y^2 + D\eta)[\hdots \Sigma_{\phi}^2] + kk_uT_{\alpha}(z+s)(z^2 - s^2) \sigma_p^2\Sigma_{yr}^2}{\Sigma_{yr}^2[\hdots \Sigma_{\phi}^2]}, \\
  \frac{B_x^2}{4A_x} &=& \eta^2 \, \frac{D^2[\hdots \Sigma_{\phi}^2]}{2\Sigma_{yr}^2[BD]} + \eta \, \frac{iDk_uT_{\alpha}(z-s)\sigma_y^2[\hdots \Sigma_{\phi}^2] + Dkk_uT_{\alpha}(z-s)(z+s)^2\sigma_p^2\Sigma_{yr}^2}{\Sigma_{yr}^2[BD]} \nonumber \\ && \quad + k_u^2T_{\alpha}^2(z-s)^2 \, \frac{k^2(z+s)^4\sigma_p^4\Sigma_{yr}^4 + 2ik[\hdots \Sigma_{\phi}^2](z+s)^2\sigma_p^2\sigma_y^2\Sigma_{yr}^2- [\hdots \Sigma_{\phi}^2]\sigma_y^4}{2\Sigma_{yr}^2[BD][\hdots \Sigma_{\phi}^2]}.
\end{eqnarray}
\subsection{The $d\eta$ integral}
We have
\begin{eqnarray}
  && \int d\eta \exp \left[ -\eta^2 \left( \frac{D^2}{2\sigma_y^2} + \frac{1}{2\sigma_{\eta}^2} - \frac{D^2\sigma_r^2}{2\sigma_y^2\Sigma_{yr}^2} - \frac{D^2[\hdots \Sigma_{\phi}^2]}{2\Sigma_{yr}^2[BD]} \right) \right. \nonumber \\ && \quad \left.+ \eta \left( -2ik_u(z-s) + \frac{iDk_uT_{\alpha}(z-s)\sigma_r^2}{\Sigma_{yr}^2} + iDk_uT_{\alpha}(z-s) \, \frac{\sigma_y^2[\hdots \Sigma_{\phi}^2] - ik(z+s)^2 \sigma_p^2\Sigma_{yr}^2}{\Sigma_{yr}^2[BD]} \right) \right],
\end{eqnarray}
whence
\begin{eqnarray}
  A_{\eta} &=& \frac{[BD]_y}{2\sigma_{\eta}^2[BD]}, \\
  B_{\eta} &=& - \frac{2ik_u(z-s)[BD] - iDk_uT_{\alpha}(z-s) \big\{ [BD]_{\sigma_r} - ik(z+s)^2\sigma_p^2 \big\}}{[BD]}, \\
  \frac{B_{\eta}^2}{4A_{\eta}} &=& - \frac{2[BD]k_u^2(z-s)^2\sigma_{\eta}^2}{[BD]_y} + \frac{2Dk_u^2T_{\alpha}(z-s)^2\sigma_{\eta}^2\big\{ [BD]_{\sigma_r} - ik(z+s)^2\sigma_p^2 \big\}}{[BD]_y} \nonumber \\ && \quad - \frac{(k_uT_{\alpha}(z-s)D\sigma_{\eta})^2\big\{ [BD]_{\sigma_r} - ik(z+s)^2\sigma_p^2 \big\}}{2[BD]_y[BD]}.
\end{eqnarray}

\subsection{Consolidating the prefactor}

The prefactor coming out of all the Gaussian integrals is
\begin{equation}
  \frac{\pi^{5/2}}{\sqrt{A_pA_{\phi}A_yA_xA_{\eta}}} = \frac{(2\pi)^{5/2}\sigma_{\phi}\sigma_p\sigma_y\sigma_r\sigma_{\eta}(z-s)}{[BD]_y}.
\end{equation}
Combining this with the gain convolution integral prefactor gives the final prefactor, after also accounting for the non-TGU $x$ dimension.

\subsection{Simplifying the exponential}

Consolidating all the surviving terms in the exponential results in the following exponent:
\begin{eqnarray}
    &&\underbrace{i k_u \Delta \nu (z-s)}_a
                       - k_u^2T_{\alpha}^2(z-s)^2 \Bigg\{ \underbrace{\frac{\sigma_{r_1}^2\sigma_{x_1}^2}{2\Sigma_{yr}^2}}_b
                       + \underbrace{\frac{\sigma_{p_1}^2(z+s)^2[\hdots \sigma_{\phi_1}^2]}{8 [\hdots \Sigma_{\phi_1}]}}_c  \Bigg. \nonumber \\ && \qquad \Bigg. - \underbrace{\frac{\left[ k(z+s)^2\sigma_{p_1}^2\Sigma_{yr}^2 + i \sigma_{x_1}^2[\hdots \Sigma_{\phi_1}^2] \right]^2}{2 \Sigma_{yr}^2[BD][\hdots \Sigma_{\phi_1}^2]}}_d + \underbrace{\frac{D^2\sigma_{\eta}^2 \left[ [BD]_{\sigma_{r_1}} - ik(z+s)^2\sigma_{p_1}^2 \right]^2}{2[BD]_y[BD]}}_e \Bigg\} \nonumber \\ && \qquad - \underbrace{\frac{2[BD]k_u^2 (z-s)^2 \sigma_{\eta}^2}{[BD]_y}}_f + \underbrace{\frac{2D k_u^2 T_{\alpha}(z-s)^2\sigma_{\eta}^2\left[ [BD]_{\sigma_{r_1}} - ik(z+s)^2\sigma_{p_1}^2 \right]}{[BD]_y}}_g,
\end{eqnarray}
where we labelled each term with lowercase letters for easy reference. Expand term $d$ to get
\begin{equation}
  d = -\underbrace{\frac{k^2(z+s)^4\sigma_p^4\Sigma_{yr}^2}{2[BD][\hdots \Sigma_{\phi}^2]}}_h - \underbrace{\frac{ik(z+s)^2\sigma_p^2\sigma_y^2}{[BD]}}_i + \underbrace{\frac{[\hdots \Sigma_{\phi}^2]\sigma_y^4}{2\Sigma_{yr}^2[BD]}}_j.
\end{equation}
Expand term $e$ to get
\begin{equation}
  e = \underbrace{\frac{D^2\sigma_{\eta}^2[BD]_{\sigma_r}^2}{2[BD]_y[BD]}}_k - \underbrace{\frac{ik D^2\sigma_{\eta}^2(z+s)^2\sigma_p^2[BD]_{\sigma_r}}{[BD]_y[BD]}}_l - \underbrace{\frac{k^2D^2\sigma_{\eta}^2(z+s)^4\sigma_p^4}{2[BD]_y[BD]}}_m.
\end{equation}
Combine the terms $c$, $h$ and $m$ to get
\begin{equation}
  c + h + m = \frac{\sigma_p^2(z+s)^2[BD]_{\sigma_{\phi}}}{8[BD]_y}.
\end{equation}
Next, combine the following terms:
\begin{equation}
  i + l = - \frac{ik(z+s)^2\sigma_p^2\Sigma_{y\eta}^2}{[BD]_y}, \qquad
  b + j + k = \frac{[BD]_{\sigma_r}\Sigma_{y\eta}^2}{2[BD]_y}.
\end{equation}
Consolidate all the terms so far to obtain
\begin{eqnarray}
  &&-ik_u\Delta \nu(z-s) - k_u^2T_{\alpha}^2(z-s)^2 \Bigg\{  \underbrace{\frac{\Sigma_{y\eta}^2[BD]_{\sigma_r}}{2[BD]_y}}_p + \underbrace{\frac{\sigma_p^2(z+s)^2[BD]_{\sigma_{\phi}}}{8[BD]_y}}_q - \underbrace{\frac{ik(z+s)^2\sigma_p^2\Sigma_{y\eta}^2}{[BD]_y}}_r
    \Bigg\} \nonumber \\ && \qquad - \underbrace{\frac{2[BD]k_u^2 (z-s)^2 \sigma_{\eta}^2}{[BD]_y}}_s + \underbrace{\frac{2D k_u^2 T_{\alpha}(z-s)^2 [BD]_{\sigma_{r}} \sigma_{\eta}^2}{[BD]_y}}_t - \underbrace{\frac{2iDk k_u^2 T_{\alpha}(z^2-s^2)^2\sigma_{\eta}^2\sigma_{p}^2}{[BD]_y}}_u.
\end{eqnarray}
Combine terms $q, r,$ and $u$ together, and terms $p,s,$ and $t$ together. This results in
\begin{eqnarray}
  &&-ik_u\Delta \nu(z-s) - \frac{k_u^2(z-s)^2}{2} \left\{ \frac{4\sigma_{\eta}^2[BD] - [BD]_{\sigma_r}( T_{\alpha}^2 \Sigma_{y\eta}^2 - 4T_{\alpha}D \sigma_{\eta}^2  )}{[BD]_y} \right\} \nonumber \\ && \quad - \frac{k_u^2\sigma_p^2(z^2-s^2)^2T_{\alpha}}{8} \left\{ \frac{T_{\alpha}[BD]_{\sigma_{\phi}} + 8ik(T_{\alpha}\Sigma_{y\eta}^2 - 2 D\sigma_{\eta}^2)}{[BD]_y} \right\}
\end{eqnarray}
Now impose the dispersion-gradient relationship:
\begin{equation}
  \alpha D = \frac{2 + K_0^2}{K_0^2} \frac{D^2\sigma_{\eta}^2}{\Sigma_{y\eta}^2}, \quad \Rightarrow \quad T_{\alpha} = \frac{2D\sigma_{\eta}^2}{\Sigma_{y\eta}^2}.
\end{equation}
Plugging that in, we obtain
\begin{equation}
  -ik_u\Delta \nu(z-s) - \frac{2k_u^2(z-s)^2\sigma_y^2\sigma_{\eta}^2}{\Sigma_{y\eta}^2} - \frac{k_u^2\sigma_p^2(z^2-s^2)^2D^2\sigma_{\eta}^4}{2\Sigma_{y\eta}^4} \frac{[BD]_{\sigma_{\phi}}}{[BD]_y}.
\end{equation}
At this point, we substitute in the definitions Eqs.~\eqref{eq:detuning} --~\eqref{eq:Sigmaphixy} to arrive at the the 3D gain formula Eq.~\eqref{eq:gain}.
\end{widetext}

\end{document}